\documentclass[aps,prd,superscriptaddress,showpacs,twocolumn]{revtex4-1}
\usepackage{graphicx}
\usepackage[caption=false]{subfig}
\usepackage{epstopdf}
\usepackage{amsmath}
\usepackage{amsfonts} 
\usepackage{amssymb}
\usepackage{latexsym}
\usepackage{hyperref}
\usepackage[english]{babel}
\usepackage[utf8]{inputenc}
\usepackage[colorinlistoftodos]{todonotes}
\usepackage{xcolor}
\usepackage{slashed}
\usepackage{feynmp}

\usepackage{bbold}
\usepackage{eufrak}

\begin{document}
\title{Lorentz violation in simple QED processes}

\author{G.P. de Brito}\email{gpbrito@cbpf.br}
\affiliation{Centro Brasileiro de Pesquisas F\'{i}sicas (CBPF), Rua Dr. Xavier Sigaud 150, Urca, Rio de Janeiro, Brazil, CEP 22290-180}

\author{J.T. Guaitolini Junior}\email{jguaitolini@ifes.edu.br}
\affiliation{Centro Brasileiro de Pesquisas F\'{i}sicas (CBPF), Rua Dr. Xavier Sigaud 150, Urca, Rio de Janeiro, Brazil, CEP 22290-180}
\affiliation{Instituto Federal do Esp\'{i}rito Santo (IFES), Av. Vit\'{o}ria 1729, Jucutuquara, Vit\'{o}ria, ES, Brazil, CEP 29040-780}

\author{D. Kroff}\email{kroff@ift.unesp.br}
\affiliation{Instituto de F\'{i}sica Te\'{o}rica, Universidade Estadual Paulista, Rua Dr. Bento Teobaldo Ferraz,
271 - Bloco II, São Paulo, SP, Brazil, CEP 01140-070}

\author{P.C. Malta}\email{malta@thphys.uni-heidelberg.de}
\affiliation{Centro Brasileiro de Pesquisas F\'{i}sicas (CBPF), Rua Dr. Xavier Sigaud 150, Urca, Rio de Janeiro, Brazil, CEP 22290-180}
\affiliation{Institut f\"ur theoretische Physik, Universit\"at Heidelberg, Philosophenweg 16, 69120 Heidelberg, Germany}

\author{C. Marques}\email{celiom@ifes.edu.br}
\affiliation{Centro Brasileiro de Pesquisas F\'{i}sicas (CBPF), Rua Dr. Xavier Sigaud 150, Urca, Rio de Janeiro, Brazil, CEP 22290-180}
\affiliation{Instituto Federal do Esp\'{i}rito Santo (IFES), Av. Vit\'{o}ria 1729, Jucutuquara, Vit\'{o}ria, ES, Brazil, CEP 29040-780}

\begin{abstract}
We determine the effect of a CPT-even and Lorentz violating non-minimal coupling on the differential cross sections for some of the most important tree-level processes in QED, namely, Compton and Bhabha scatterings, as well as electron-positron annihilation. Experimental limits constraining the allowed deviation of the differential cross sections relative to pure QED allow us to place upper bounds on the Lorentz violating parameters. A constraint based on the decay rate of para-positronium is also obtained. 
\end{abstract}

\maketitle


\section{Introduction} \label{sec_intro}

\indent

The Standard Model of elementary particles, based upon gauge and Lorentz symmetries, has been very successful in the last decades, with impressive experimental confirmation of most of its theoretical predictions \cite{PDG}. Nevertheless, it cannot be a final theory, but only a low energy effective limit of some more fundamental high energy theory. In some scenarios beyond the Standard Model, e.g. string theory, Lorentz symmetry -- and eventually CPT symmetry -- is violated \cite{Kost1, Posp1, Mavro1, Mavro2}, thereby generating low energy effective interactions. This terms are generally suppressed by the inverse power of some large energy scale \cite{Colladay} and could give rise to small, though interesting, effects in the physics at scales accessible today or in the near future.

During the last two and a half decades, Lorentz symmetry violation has been widely studied both from the theoretical and experimental points of view. On the theoretical side, a comprehensive list of possible Lorentz-violating (LV) terms has been developed, what constitutes the so-called Standard Model Extension (SME) \cite{Colladay,Colladay2}. The SME complements the usual Standard Model by including a plethora of novel LV interactions in all its sectors. On the experimental side, tests in several areas over different energy scales, ranging from atomic spectroscopy to astrophysics, have placed very strong bounds on the possible LV coefficients -- see e.g. \cite{Table, Mattingly} and, more recently, \cite{EXO}.

An interesting way to introduce Lorentz violation in the otherwise Lorentz-preserving QED is to modify the electron-photon vertex directly. This can be done in a gauge-invariant way by coupling a constant, i.e., space-time-independent, 4-vector background, $\xi^{\mu}$, with the usual field-strength tensor, $F_{\mu\nu}$. This derivative coupling would therefore modify the standard Lagrangian, which now reads
\begin{equation} 
\mathcal{L} = -\frac{1}{4}F_{\mu\nu}^2 + \bar{\psi}(i \gamma^\mu \partial_\mu - e A_\mu \gamma^\mu - m) \psi + 
\xi^{\mu}\bar{\psi}\gamma^{\nu}\psi F_{\mu\nu}, \label{case_1}
\end{equation}
where $m$ and $e$ are the electron's mass and electric charge. We omit the gauge-fixing term here, since we are dealing with conserved external currents.

The LV background $\xi$ is a non-minimal coupling with canonical dimensions of inverse mass. This LV scenario has been proposed in ref.\cite{bakke} in the context of topological phases and represents a very simple gauge-invariant non-minimal coupling possibility (it is important to note that this coupling is not considered in the SME  \cite{Colladay,Colladay2}). Given that $\xi$ is fixed, it plays the role of a non-dynamical background and Lorentz symmetry is broken, as it selects a preferred direction in space-time.

It is easy to see that this LV interaction acts as a non-minimal coupling changing the usual covariant derivative to $D_{\mu} = \partial_{\mu} + ieA_{\mu} - i\xi^{\nu}F_{\mu\nu}$, whereby the extra term is clearly gauge invariant. This CPT-even modification affects all electron-photon interactions already at tree level and similar derivative non-minimal couplings have also been proposed in several instances: quantum mechanics and the hydrogen atom \cite{belich}, magnetic and electric dipole moments \cite{pospelov,pospelov2} -- also as an interesting way to generate a tree-level magnetic moment for Majorana neutrinos \cite{belich2} -- and scattering processes \cite{Maluf}.


Here we will focus on the latter case and discuss the impact of the last term in eq.\eqref{case_1} in a few simple and well-know QED reactions, namely: Compton and Bhabha scatterings, electron-positron annihilation and the life-time of para-positronium. For simplicity, in the following we shall only keep terms in the squared amplitudes, and consequently also in the differential cross sections and decay rates, up to leading order in the LV parameter. This is a good level of approximation, once LV effects have not been conclusively observed, so it is expected that the associated parameters are very small. Also, as we consider a dimension-5 operator, restricting our analysis to tree level processes, we can ignore renormalizability-related issues.

Scattering processes have also been considered in detail in ref. \cite{Colladay3}, where the authors find that, due to modifications in the propagators (already at tree level), linear momentum and velocity may be misaligned, therefore making the task of computing cross sections a bit trickier. In the present work we evade this issue, as the bare propagators are left intact -- possible modifications may arise at the quantum level, though. This issue is very interesting, but lies outside of the scope of this paper.

As mentioned above, $\xi$ is a constant 4-vector playing the role of a fixed background, and it can be decomposed in spherical coordinates, with the axes adequately chosen according to the process of interest. The polar ($\theta_{\xi}$) and azimuthal ($\phi_{\xi}$) angles are fixed relative to the experimental set-up at a given time and, as will become clear in the following, the extra momentum factor introduced by the LV non-minimal coupling will produce terms proportional to $\xi\cdot p_i$, where $p_i$ are the momenta of the in-coming and out-going particles. These momentum-dependent terms will be responsible for new angle and energy profiles for the respective reactions, especially regarding the azimuthal angle, so that these non-standard features act as LV signatures to be searched for experimentally.

\begin{figure}[htb!]
\centering
\includegraphics[width=0.8\linewidth]{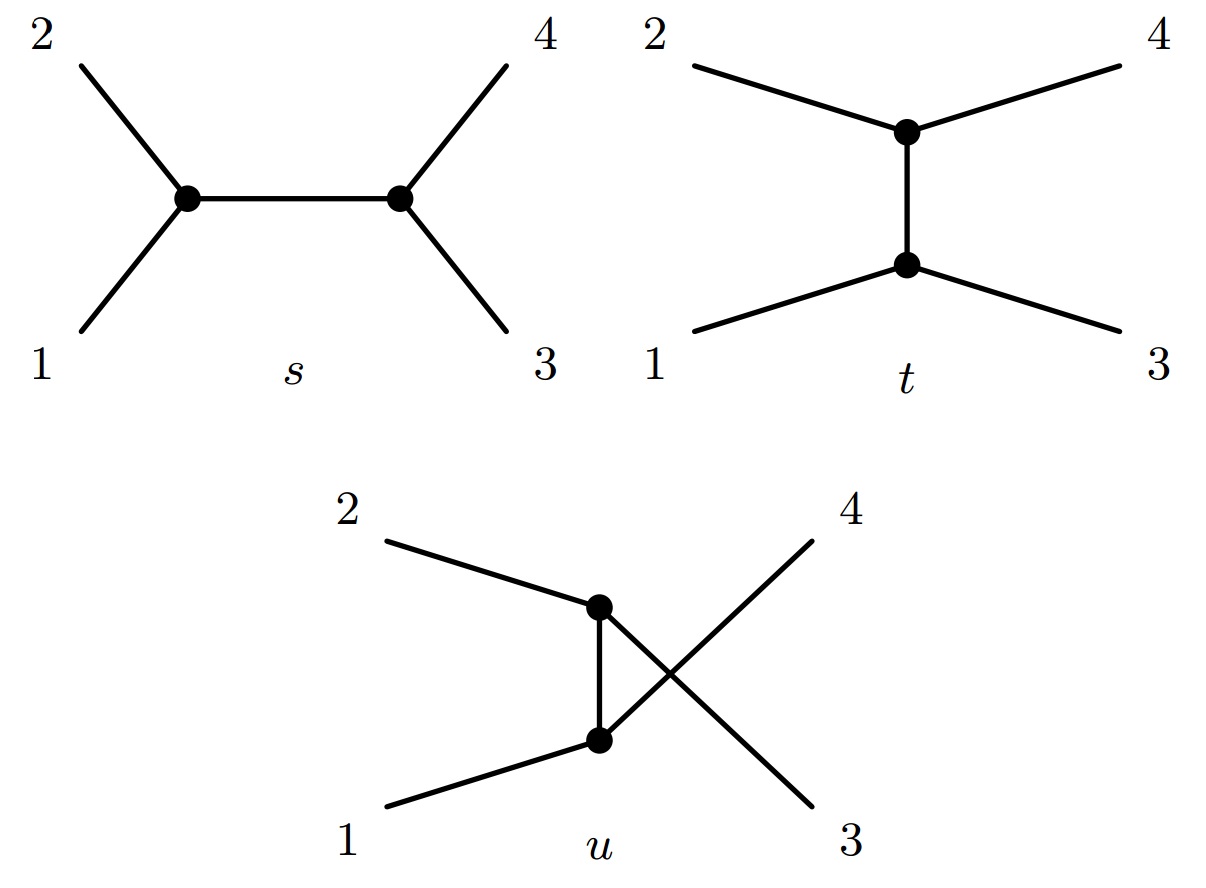} 				
\caption{Feynman diagrams contributing to the processes considered and associated Mandelstam variables. The blob represents the effective vertex, eq.\eqref{vertex}.}
\label{fig:diagrams}
\end{figure}

The LV-modified Lagrangian, eq.\eqref{case_1}, translates into an extension of the usual QED $ee\gamma$ vertex, namely
\begin{equation}
i \Gamma^\mu = i e \gamma^\mu + \slashed{q} \, \xi^\mu - (\xi \cdot q) \gamma^\mu, \label{vertex}
\end{equation}
with $q$ here representing the 4-momentum carried by the photon line, conventioned as being positive (negative) for in-coming (out-going)
photons. In fig.\eqref{fig:diagrams} below we present the generic $s$-, $t$- and $u$-channel tree-level Feynman diagrams that contribute to the processes we consider. The blob indicates the modified vertex, eq.\eqref{vertex}, and for Bhabha and Compton scatterings only the $s$- and $t$-channels play a role, whereas for electron-positron annihilation only the $t$- and $u$-channels contribute.

Our goal is to obtain the modifications brought up by the LV piece of the new vertex and, through experimental limits on deviations from the Lorentz preserving QED, establish upper bounds on the components of $\xi$. The LV parameters associated with other similar derivative non-minimal couplings (e.g. involving the dual field-strength tensor) have been constrained to be $\lesssim 10^{-3} \, {\rm GeV}^{-1}$ \cite{Maluf}, and we shall extract limits of similar magnitude from Bhabha scattering and unpolarized electron-positron annihilation, while the bounds from para-positronium are somewhat looser.

A practical comment is in order: the background $\xi$ is assumed to be completely non-dynamical, i.e., it is fixed in space-time. This is truly explicit only in an (approximately) inertial reference frame, e.g. the Sun-centered frame (SCF) \cite{ref_sun}, so that, in comparison, Earth is not ``inertial enough" due to its sideral and orbital motions. Consequently, in Earth-bound experiments, also in satellites \cite{sat}, the LV-modified observables (e.g. cross sections) should display daily and/or yearly time-modulations. Since experiments are not performed entirely during small fractions of a single day, but over periods of days spread over months, the experimental signals would effectively give information on the time-averaged LV parameters.

To discuss the precise time-dependence it is necessary to express the LAB-frame components, $\xi^{\mu} = \xi^{\mu}_{\rm lab}$, in terms of the components, $\xi_{\rm sun}^{\mu}$, which are fixed (static) in the Sun-centered frame. Following refs.\cite{ref_sun,Table}, we find that, up to first order in $\beta$ (Earth's orbital velocity $\sim 10^{-4}$), $\xi_{\rm lab}^{0} = \xi_{\rm sun}^{T} + \mathcal{O}(\beta)$ and $\xi_{\rm lab}^{i} = R^{iJ} (\chi,T) \xi_{\rm sun}^{J} + \mathcal{O}(\beta)$, where $R^{iJ} (\chi,T)$ is the rotation matrix depending on the time $T$ in the Sun-centered frame and co-latitude $\chi$ of the experiment on Earth. The above expressions clearly display the time-dependent nature of the background 4-vector -- specially its spatial part -- in the Earth-bound reference frame.

In the context of the approximations above, the time components are easily factored out from the squared amplitudes. This is not so simple for the spatial components, which involve a time-dependent rotation matrix. As one of our goals is to place bounds on the LV parameters, we necessarily need to compare our results with the experimental ones. Also, due to the already mentioned non-inertial character of the Earth-bound reference frame, we must include the time-averaging procedure in our computations. This is a trivial task for the time components, as -- within our approximations -- they are identical in both frames.

On the other hand, the space components always come together with the time-average of some rotation, what, allied to the fact that our bounds are simply estimates coming from the experimental uncertainties, yields a very cumbersome expression which is not easily readable or even converted into a bound. Therefore we shall only explicitly present the limits on the time components of the background. The corresponding limits on combinations of the space components should be about the same order as the ones for the time components, as we do not expect the averaging procedure to introduce strong suppressing factors.

This paper is organized as follows: in section \ref{sec_compton} and \ref{sec_bhabha} we present and discuss the LV-modified Compton and Bhabha scatterings, respectively, while in section \ref{sec_ee} we 
analyse the unpolarized electron-positron annihilation in two photons. We also used the life-time of para-positronium to find an upper limit on the LV parameters. In section \ref{sec_conclusion} we 
present our concluding remarks. In our calculations we have used the Package-X \cite{PackageX} to automatically evaluate the traces and contractions from the averaging procedure.



\section{Compton scattering} \label{sec_compton}

\indent

As a preliminary examination of the effect of the modified vertex, eq.\eqref{vertex}, we consider Compton scattering, i.e., the process by which radiation of energy $\omega$ is scattered by a free electron, usually assumed at rest. The final out-going photons are emitted with a different energy, $\omega' \leq \omega$, at an angle $\theta$ and, due to 4-momentum conservation, we find that the initial and final photon energies are related via $\frac{\omega'}{\omega} = \left[ 1 + \frac{\omega}{m}(1 - \cos\theta)  \right]^{-1}$, the well-known Compton frequency shift result.

In QED, the electron-photon scattering, $e^{-}(p_1) \, + \, \gamma(p_2) \rightarrow \gamma(p_3) \, + \, e^{-}(p_4)$, can be represented by the $s$- and $t$-channel Feynman diagrams indicated in fig.\ref{fig:diagrams}. This process was first studied by Klein and Nishima \cite{KN, Mandl_Shaw} and was one of the first applications of the then new Dirac quantum mechanics. The differential cross section reads:
\begin{equation}
\frac{d\sigma^{e\gamma}_{_{QED}}}{d\Omega} = \frac{\alpha^2}{2m^2}\left( \frac{\omega'}{\omega} \right)^2 \left[ \frac{\omega'}{\omega} + \frac{\omega}{\omega'} - \sin^2\theta  \right],  \label{KN_qed}
\end{equation}
whose low energy limit reproduces the classical Thomson scattering differential cross section $\sim (1 + \cos^2\theta)$, which is energy-independent. In this section we will focus on two opposite energy regimes: $\omega \ll m$ and $\omega \gg m$. In the former, the process is elastic, i.e., $\omega' \simeq \omega$, while in the latter the Compton formula gives $\omega' \simeq m(1 - \cos\theta)^{-1}$.

We are interested in determining the effect of $i\Gamma^{\mu}$ to the scattering of radiation off static free electrons. The total amplitude for this process may be decomposed in two pieces, a pure QED piece and a LV one, $\mathcal{M}_{\rm tot} = \mathcal{M}_{_{\rm QED}} + \mathcal{M}_{\xi}$. Here we will work with unpolarized electrons and photons in the LAB system, where the electron is initially at rest. In this particular reference frame the participating particles have momenta given by $p_1 = m(1,0,0,0)$, $p_2 = \omega(1,0,0,1)$, $p_3 = \omega'(1,\sin\theta\cos\phi,\sin\theta\sin\phi,\cos\theta)$, with $p_4$ fixed by 4-momentum conservation.


The deviation from the Klein-Nishima result is determined by $\langle |\mathcal{M}_{_{\rm LV}}|^2 \rangle = \langle \mathcal{M}_{_{\rm QED}}\mathcal{M}_{\xi}^{\ast} \rangle + \langle \mathcal{M}_{_{\rm QED}}^{\ast}\mathcal{M}_{\xi}  \rangle + \langle |\mathcal{M}_{\xi}|^2 \rangle$, where $\mathcal{M}_{_{\rm LV}}$ contains, in principle, terms of first and second order in the background. We are not going to provide the explicit lengthy expression for the squared amplitude here, but we would like to remark, however, that part of the QED-LV interference terms -- the one of $\mathcal{O}(\xi)$ -- turns out to be purely imaginary, thus canceling automatically. The LV modified Klein-Nishima formula is then given by
\begin{equation}
\begin{split}
\frac{d\sigma^{e\gamma}_{_{\rm LV}}}{d\Omega} & = \frac{\alpha}{8\pi m^2} [(\xi \cdot p_2)^2 + (\xi \cdot p_3)^2] \left( \frac{\omega'}{\omega} \right)^2 \times \\
&\times \left[ \frac{\omega'}{\omega} + \frac{\omega}{\omega'} - \sin^2\theta  \right].  \label{KN_lv}
\end{split}
\end{equation}
Interestingly enough, this result shares great similarity with its Lorentz-preserving counterpart, eq.\eqref{KN_qed}.

An important observation is that our LV result is of first order in $\alpha$, as opposed to QED, which is of second order in the fine structure constant. Also, following the momenta assignments given above, $(\xi \cdot p_2)^2 \simeq \omega^2$ and $(\xi \cdot p_3)^2 \simeq \omega'^2$, so that the LV differential cross section, eq.\eqref{KN_lv}, can be distinguished from the standard Klein-Nishima formula not only by its angular dependence, but also through its energy-dependent profile.

Before we proceed, we would like to comment on the general energy behavior of eq.\eqref{KN_lv}. If we define $x=\omega/m$ and $\mathcal{P}(x,\theta) = 1 + x(1 - \cos\theta)$, we may rewrite the LV-modified Klein-Nishima formula as
\begin{equation}
\begin{split}
\frac{d\sigma^{e\gamma}_{_{\rm LV}}}{d\Omega} &= \kappa \tilde{\xi}^2 \frac{x^2}{\mathcal{P}(x,\theta)^2} \left[ 1 + \mathcal{P}(x,\theta)^{-2} \right] \times \\
&\times \left[ \mathcal{P}(x,\theta) + \frac{1}{\mathcal{P}(x,\theta)} - \sin^2\theta \right],
\end{split}
\end{equation}
with $\kappa = \frac{\alpha}{8\pi}$ and $\tilde{\xi}^2$ containing the dimensionless angular factors from $(\xi \cdot p_2)^2 + (\xi \cdot p_3)^2$. The extra $x^2$ energy factor in the numerator owes its presence to the electromagnetic field-strength tensor in the LV non-minimal coupling.

For low frequency incident radiation, $x \ll 1$, we notice that $\mathcal{P}(x,\theta) \rightarrow 1$, so that, apart from trigonometric functions, $\frac{d\sigma^{e\gamma}_{_{\rm LV}}}{d\Omega} \sim x^2$. This means that, relative to the standard Thomson result, the low energy limit of the LV differential cross section is generally strongly suppressed, thus compromising any hope of experimental verification in this energy regime.

On the other extreme of the spectrum, for high frequencies, $x \gg 1$, we have $\mathcal{P}(x,\theta) \rightarrow x$, modulo angular factors, so that $\mathcal{P}(x,\theta)^{-1} \sim 0$ while $x\mathcal{P}(x,\theta)^{-1} \sim 1$. With this we find a linear energy dependence, $\frac{d\sigma^{e\gamma}_{_{\rm LV}} }{d\Omega} \sim x$, and we conclude that the LV-induced modifications are actually amplified in the high energy regime. It is worthwhile mentioning that the corresponding limit of the usual Klein-Nishima formula (for not too small scattering angles) is found to be $\frac{d\sigma^{e\gamma}_{_{\rm QED}} }{d\Omega} \sim x^{-1}$, i.e., classically the electron is not a good scattering target for highly energetic incident photons. This is clearly contrasting with our LV results, whose signal may be optimally distinguished from those of standard QED at increasing energies.

As stated above, we are interested in determining the angular profiles emerging in the low and high energy limits, so let us start with the first, where we may assume that the electron is a fixed target and the photon bounces off elastically, i.e., $\omega' \simeq \omega$. To proceed we need to specify the nature of the background and evaluate eq.\eqref{KN_lv} accordingly, so we choose to start with $\xi^{\mu} = (\xi^0,0)$. In this scenario all angular information contained in $\xi \cdot p_2$ and $\xi \cdot p_3$ is lost and we have $(\xi \cdot p_2)^2 + (\xi \cdot p_3)^2 \rightarrow 2\xi_0^2\omega^2$, so that the LV differential cross section becomes
\begin{equation}
\frac{d\sigma^{e\gamma, \, \xi_0}_{_{\rm LV}}}{d\Omega} = \frac{\alpha \xi_0^2}{4\pi} \left( \frac{\omega}{m} \right)^2 \left[ 1 + \cos^2\theta  \right],  \label{lv_0}
\end{equation}
whose angular profile is the same as in the classical Thomson result. As discussed above, numerically, however, this differential cross section is heavily suppressed relative to the QED one not only via the small coupling constant, but also through the extra $(\omega/m)^2$ factor.   


We consider next the case of a pure space-like background, for which we expect stronger angular dependence in comparison to the QED Klein-Nishima formula. For the sake of simplicity, we content ourselves with two physically interesting scenarios, namely, ${\boldsymbol{\xi} } \parallel {\bf \hat{z}}$ and $\boldsymbol{\xi} \perp {\bf \hat{z}}$. The first case corresponds to a background aligned with the direction of propagation of the incident photon, while the second is lying in the transverse plane.

\begin{figure}[htb]
\includegraphics[width=.85\linewidth]{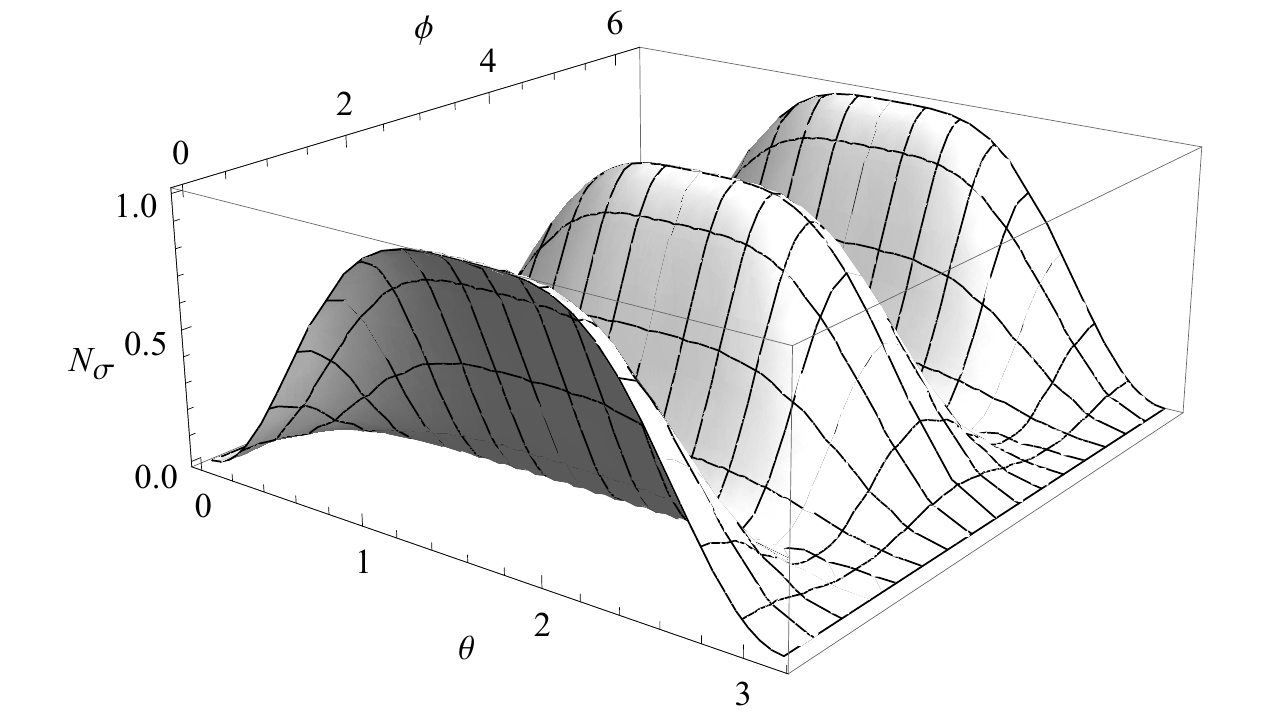}

\vspace{0.5cm}
\includegraphics[width=.8\linewidth]{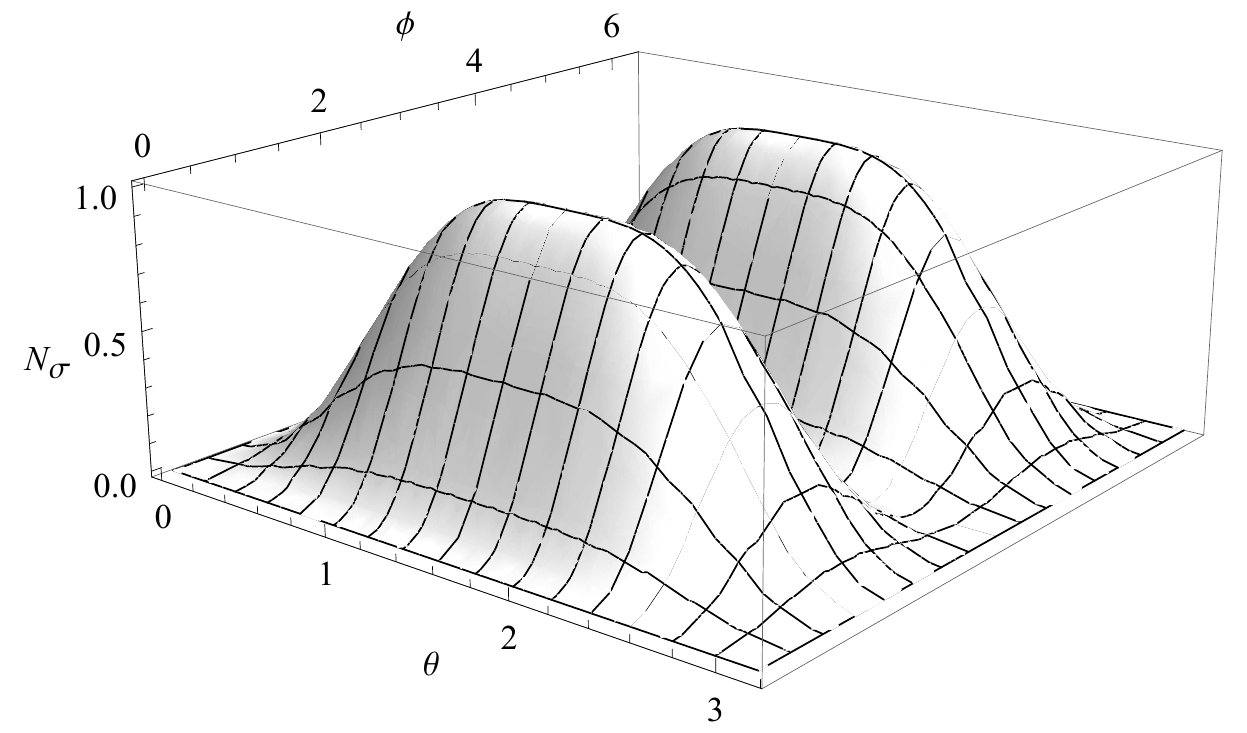}
\caption{Instantaneous angular profile (low-energy regime) of eq.\eqref{lv_trans} for $\phi_\xi = 0$ (top) and $\phi_\xi = \pi/2$ (bottom), with $N_{\sigma} = \left[ \frac{\alpha |\boldsymbol{\xi}|^2}{8\pi} \left( \frac{\omega}{m} \right)^2 \right]^{-1} d\sigma^{e\gamma, \, \perp}_{_{\rm LV}}/d\Omega$.}
\label{fig1}
\end{figure}

For a background parallel to the $z$-axis ($\theta_\xi = 0$) there is no azimuthal dependence, but $(\xi \cdot p_2)^2 + (\xi \cdot p_3)^2$ introduces an additional $(1+\cos^2\theta)$ factor, so that
\begin{equation}
\frac{d\sigma^{e\gamma, \, \parallel}_{_{\rm LV}}}{d\Omega} = \frac{\alpha |\boldsymbol{\xi}|^2}{8\pi} \left( \frac{\omega}{m} \right)^2 \left[ 1 + \cos^2\theta  \right]^2,
\end{equation}
whose angular dependence is steeper than that of the pure time-like case. More interesting is the second scenario -- a transverse background -- with $(\xi \cdot p_2)^2 = 0$ and $(\xi \cdot p_3)^2 = |\boldsymbol{\xi}|^2\omega^2\sin^2\theta\cos^2(\phi - \phi_\xi)$, showing that, in these circumstances, a distinctive azimuthal signature appears. The corresponding differential cross section is
\begin{equation}
\frac{d\sigma^{e\gamma, \, \perp}_{_{\rm LV}}}{d\Omega} \!=\! \frac{\alpha |\boldsymbol{\xi}|^2}{8\pi} \left( \frac{\omega}{m} \right)^2 \sin^2\theta\cos^2(\phi -\! \phi_\xi)\left[ 1 \!+\! \cos^2\theta  \right],  \label{lv_trans}
\end{equation}
whose instantaneous angular profile is plotted in fig. \ref{fig1} for different relative orientations of the background in the transverse $xy$-plane. 

Now we turn to the high energy regime, $\omega \gg m$. As pointed out in the beginning of this section, the Compton formula indicates that $\omega'$ is approximately $\omega$-independent and $\left( \frac{\omega'}{\omega} \right)^2 \left[ \frac{\omega'}{\omega} + \frac{\omega}{\omega'} - \sin^2\theta  \right] \simeq \frac{m}{\omega (1 - \cos\theta)}$ in such a way that $\frac{d\sigma^{e\gamma}_{_{\rm QED}}}{d\Omega} = \frac{\alpha^2}{2m\omega} \left( 1 - \cos\theta \right)^{-1}$. As before, let us first consider $\xi^{\mu} = (\xi^0, 0)$, where $(\xi \cdot p_2)^2 \rightarrow \xi_0^2\omega^2$, but now, different from the $\omega \ll m$ case, the contribution from $(\xi \cdot p_3)^2$ is negligeable to $\mathcal{O}(m^2/\omega^2)$. The LV differential cross section for a time-like background is then 
\begin{equation}
\frac{d\sigma^{e\gamma, \, \xi_{0}}_{_{\rm LV}}}{d\Omega} = \frac{\alpha \xi_0^2}{8\pi} \frac{\omega}{m (1 - \cos\theta)}, \label{lv_00}
\end{equation}
whose angular dependence is the same as the one from usual Klein-Nishima formula.

We move next to the case of a pure space-like background and we once more focus on the particular scenarios where $\boldsymbol{\xi} \parallel {\bf \hat{z}}$ and $\boldsymbol{\xi} \perp {\bf \hat{z}}$. Proceeding in a similar fashion as in the low energy case, we find that the respective LV differential cross sections are
\begin{subequations}
\begin{equation}\label{lv_par}
\frac{d\sigma^{e\gamma, \, \parallel}_{_{\rm LV}}}{d\Omega} = \frac{\alpha |\boldsymbol{\xi}|^2}{8\pi}  \frac{\omega}{m (1 - \cos\theta)} + \mathcal{O}(m^2/\omega^2),  \, \, \, \,\ \, \text{for} \, \,\theta_\xi = 0
\end{equation}
and
\begin{equation}\label{lv_perp}
\frac{d\sigma^{e\gamma, \, \perp}_{_{\rm LV}}}{d\Omega} \sim  \mathcal{O}(m/\omega),    \, \, \, \,\ \, \text{for} \,\, \theta_\xi = \pi/2.
\end{equation}
\end{subequations}

A comment regarding the result for a transverse background is in order: eq.\eqref{lv_perp} owes its seemingly odd energy behavior to the fact that, for $\theta_\xi = \pi/2$, $\xi \cdot p_2 = 0$ while $\xi \cdot p_3 \sim \omega' \sim m$, so that no extra $\omega^2$-factor from $(\xi \cdot p_2)^2 + (\xi \cdot p_3)^2$ is available to cancel the remaining $\omega^{-1}$ from phase space. The unexpected absence of azimuthal dependence in eq.\eqref{lv_perp} -- as opposed to eq.\eqref{lv_trans} -- is not a general feature, though. For $\theta_\xi \neq \pi/2$ the distinctive $\phi$-dependent contribution is indeed recovered, albeit being strongly suppressed, since $\xi \cdot p_2 \sim \omega\cos\theta_\xi$ is no longer zero and dominates over $\xi \cdot p_3 \sim m \cos(\phi-\phi_\xi)$. In this more general situation, the aforementioned linear energy dependence of $\frac{d\sigma^{e\gamma, \, \perp}_{_{\rm LV}}}{d\Omega}$ is also expected to be re-obtained.

In ref. \cite{altschul} Compton scattering is also considered in a LV scenario -- the author works with the $\slashed{b}\gamma_5$ modification to the Dirac equation, what directly affects the electron propagator. There the differential cross section grows very rapidly for low frequencies, not recovering the classical Thomson result. This is not the case here, as we introduced a modification only in the electron-photon vertex, thus keeping the fermionic propagator -- and dispersion relation -- untouched at tree level.

Finally, we would like to note that the Compton scattering is sensitive to the refraction index of vacuum \cite{ref_ind}. Even though we do not address this point here -- it is out of the scope of this paper -- we acknowledge that this would be an interesting direction for future research, also as a means to extract limits on the LV parameters.


\section{Bhabha scattering} \label{sec_bhabha}

\indent

Bhabha scattering is the ultra-relativistic scattering of electrons and positrons and is one of the most basic and well-studied processes, serving as a high-luminosity monitor and a tool for the study of both QED and electroweak interactions \cite{LEP, SLD, Likic}. Due to its relative simplicity, Bhabha scattering has been used as a test for different beyond the Standard Model scenarios, such as theories with extra dimensions \cite{Dutta}, generalized QEDs \cite{Bufalo} and LV \cite{Maluf}.


In this section we investigate the LV-modified amplitudes for Bhabha scattering in the context of the Lagrangian from eq.\eqref{case_1}, but, before we proceed with our computations, let us briefly recall the main results from usual QED. The electron-positron scattering, $e^{-}(p_1) \, + \, e^{+}(p_2) \rightarrow e^{-}(p_3) \, + \, e^{+}(p_4)$, is usually evaluated in the center of mass (CM) frame and can be represented by the $s$- and $t$-channel Feynman diagrams depicted in fig.\eqref{fig:diagrams}.

For our purposes, we will restrict ourselves to unpolarized cross sections, hence, we have to average the squared amplitude over spins. In the CM frame the 4-momenta of the in-coming particles are $p_1 = (E,\textbf{p})$ and $p_2 = (E,-\textbf{p})$, and, for the out-going particles, $p_3 = (E,\textbf{p}')$ and $p_4 = (E,-\textbf{p}')$, with $E = E_{_{\rm CM}}/2$. For definitiveness, let us consider the initial momenta oriented along the $z$-axis, i.e., $ \textbf{p} = E \, {\bf \hat{z}}$, while the final momentum is $\textbf{p}' = E \, (\sin\theta \cos\phi, \sin\theta \sin\phi, \cos\theta)$. With these definitions, the unpolarized QED differential cross section for Bhabha scattering is
\begin{equation} \label{sec_choque_QED}
\frac{d \sigma^{ee}_{_{\rm QED}}}{d \Omega} = \frac{\alpha^2 (7 + \cos2\theta )}{16 E^2_{_{\rm CM}} ( \cos\theta - 1)^2} \, , 
\end{equation}
and in the following we shall discuss the LV modifications to this standard result.



Using $i\Gamma$ one can compute the Feynman amplitudes, which, after averaging over spins, give $\langle |\mathcal{M}_{\rm tot}|^2 \rangle = \frac{1}{4}\sum |\mathcal{M}_{t-ch} - \mathcal{M}_{s-ch}|^2$. The complete expression for $\langle |\mathcal{M}_{\rm tot}|^2 \rangle$ contains the usual QED contribution leading to eq.\eqref{sec_choque_QED} and an additional LV term, $\langle |\mathcal{M}_{_{\rm LV}}|^2 \rangle$, of order $\mathcal{O}(\xi^2)$. For the sake of simplicity, we refrain from writing $\langle |\mathcal{M}_{_{\rm LV}}|^2 \rangle$ explicitly and examine below a few particular cases of physical interest.




In order to present the differential cross section for Bhabha scattering, let us divide our analysis in the physically meaningful sub-cases of pure time- and space-like background 4-vectors. The total differential cross section is composed of the usual QED contribution, eq.\eqref{sec_choque_QED}, with an additional term coming from $\langle |\mathcal{M}_{_{\rm LV}}|^2 \rangle$. For a pure time-like background, $\xi^{\mu} = (\xi^0,0)$, we find
\begin{align}
\frac{d {\sigma}^{ee, \, \xi_0}}{d \Omega} & =\frac{\alpha^2 (7 + \cos 2\theta )}{16 E^2_{_{\rm CM}} (\cos\theta - 1)^2} \\ 
& + \frac{\alpha \, \xi_0^2   \left( \cos\theta + 2\cos^2\theta  -  \cos^3\theta  + 2 \right) \sin^2\frac{\theta}{2}}{4 \pi (\cos \theta -1)^2}, \nonumber
\label{cs_bhabha_n0}
\end{align}
where we used the aforementioned momenta attributions in the CM frame. The angular profile of this result is displayed in fig.\eqref{bhabha_fig1} below.

\begin{figure}[htb!]
\includegraphics[width=0.98\linewidth]{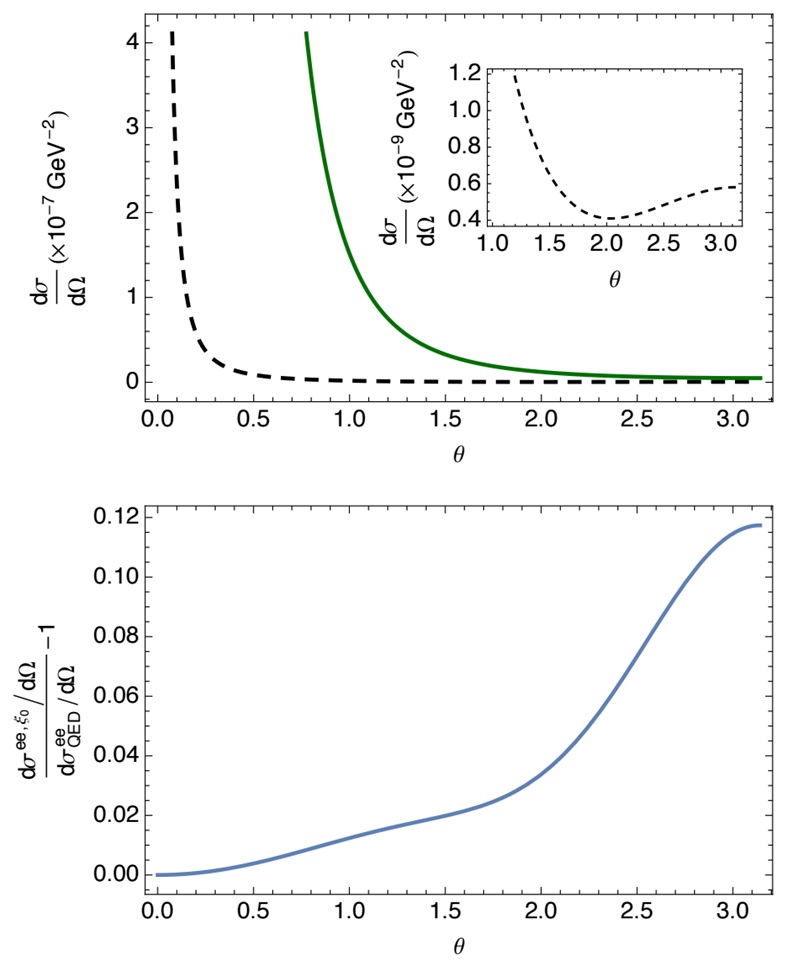}
\caption{The top panel displays the differential cross sections for the usual QED (solid line) and the time-like LV contribution (dashed line) assuming $\xi_0 = 10^{-3} \,\textmd{GeV}^{-1}$ and the inset shows the non-monotonic angular dependence of the LV sector. The bottom panel shows the deviation (LHS of eq.\eqref{criterio_1}) as a function of the scattering angle}
\label{bhabha_fig1}
\end{figure}




Small deviations from the usual tree-level results from QED for Bhabha scattering have been experimentally bounded at $\sqrt{s} = 29 \, {\rm GeV}$ and $|\cos\theta| < 0.55$ (at the PEP storage ring facility)  as \cite{Derrick}
\begin{equation}
\Bigg| \frac{d {\sigma}^{ee, \, \xi_0}/d \Omega}{d \sigma^{ee}_{_{\rm QED}}/d \Omega} - 1 \Bigg| \lesssim \frac{3 E_{_{\rm CM}}^2}{\Lambda^2}, \label{criterio_1}
\end{equation}
at the $95\%$ CL, where $\Lambda$ parametrizes possible experimental deviations from the theoretical results. The leading order contribution for the LHS of the above inequality is of order $\sim \xi_0^2  E_{_{\rm CM}}^2/\alpha^2$ and, comparing with the RHS, we arrive at $\xi_0 \lesssim \sqrt{\alpha}/\Lambda$, so that, using the limit above with $\Lambda \sim 200 \, {\rm GeV}$, we obtain the upper bound:
\begin{equation}
\xi_0  \lesssim 10^{-3} \, {\rm GeV}^{-1}, \label{limit_1}
\end{equation}
which is compatible with the results for analogous non-minimal couplings as presented in ref.\cite{Maluf}.

As one can see from the right panel in fig.\eqref{bhabha_fig1}, where the upper limit obtained above is assumed, the deviation from pure QED grows considerably with the scattering angle. We remark, however, that our estimate is consistent: the experimental limit used is valid for $|\cos \theta| < 0.55$ with the upper bound $3E_{_{\rm CM}}^2/\Lambda^2 \sim 0.06$ \cite{Derrick}. In this angular range we have $\big| \frac{d {\sigma}^{ee, \, \xi_0}/d \Omega}{d \sigma^{ee}_{_{\rm QED}}/d \Omega} - 1 \big| \sim 0.03$, cf. LHS of eq.\eqref{criterio_1} with $\xi_0 = 10^{-3} \, {\rm GeV}^{-1}$, showing that our leading order analysis is valid.


Despite of the apparently similar sizes of the LV and QED contributions in fig.\eqref{bhabha_fig1} (left panel) for large $\theta$, we see from the right panel that, for $\theta \simeq \pi$, $\big| \frac{d {\sigma}^{ee, \, \xi_0}/d \Omega}{d \sigma^{ee}_{_{\rm QED}}/d \Omega} - 1 \big|$ attains a maximum of $\sim 0.12$. This large-angle region is however hardly accessible in collision experiments and lies beyond the scope of the experimental limit used above \cite{Derrick}, i.e., $|\cos \theta| < 0.55$. For smaller values of $\xi_0$ this deviation decreases accordingly. Both panels in fig.\eqref{bhabha_fig1} suggest that measurements in the backward direction would be a promising way, though technically challenging, to look for signals of a purely time-like LV background.

Now we turn to the case of a purely spatial background 4-vector, $\xi^{\mu} = (0,\boldsymbol{\xi} )$. In this situation, the pure LV piece of the differential cross section is found to be 
\begin{eqnarray}
\frac{d {\sigma}^{ee, \, \boldsymbol{\xi}}_{_{\rm LV}}}{d \Omega} & \!= & \frac{\alpha \,|\boldsymbol{\xi}|^2  \left( 17 \cos\theta + 2\cos 2\theta - \cos 3\theta + 46 \right)}{128 \pi (\cos \theta -1)^2} \\ 
& \times & \left[\cos (\phi - \phi_\xi) \sin \theta  \sin \theta_\xi +(\cos\theta -1) \cos \theta_\xi  \right]^2, \nonumber 
\label{bhabha_n}
\end{eqnarray}
and the complete expression for the differential cross section is, as before, the combination of the formula above with the standard QED result (cf. eq.\eqref{sec_choque_QED}).

\begin{figure}[htb]
\includegraphics[width=0.8\linewidth]{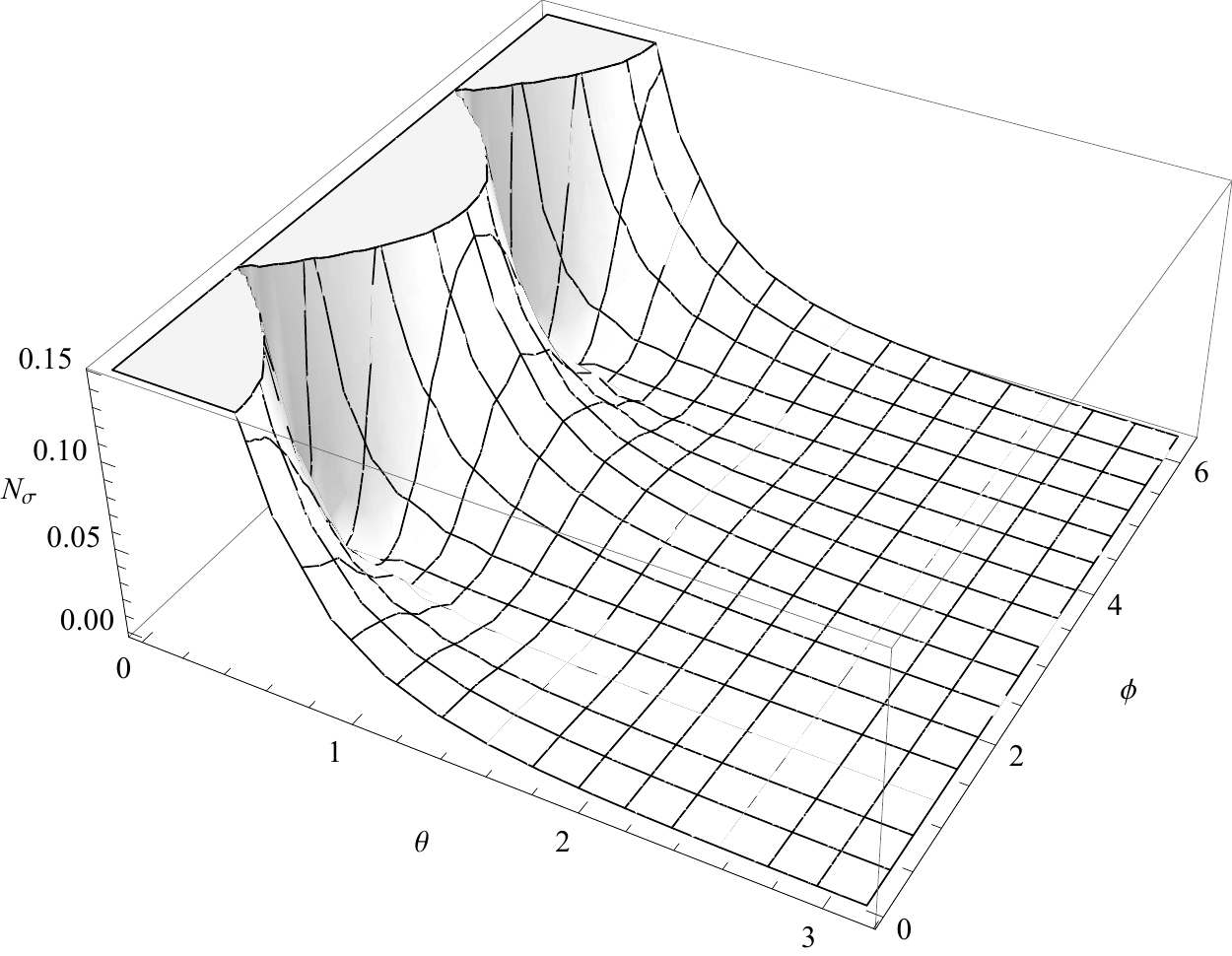}

\vspace{0.5cm}
\includegraphics[width=0.8\linewidth]{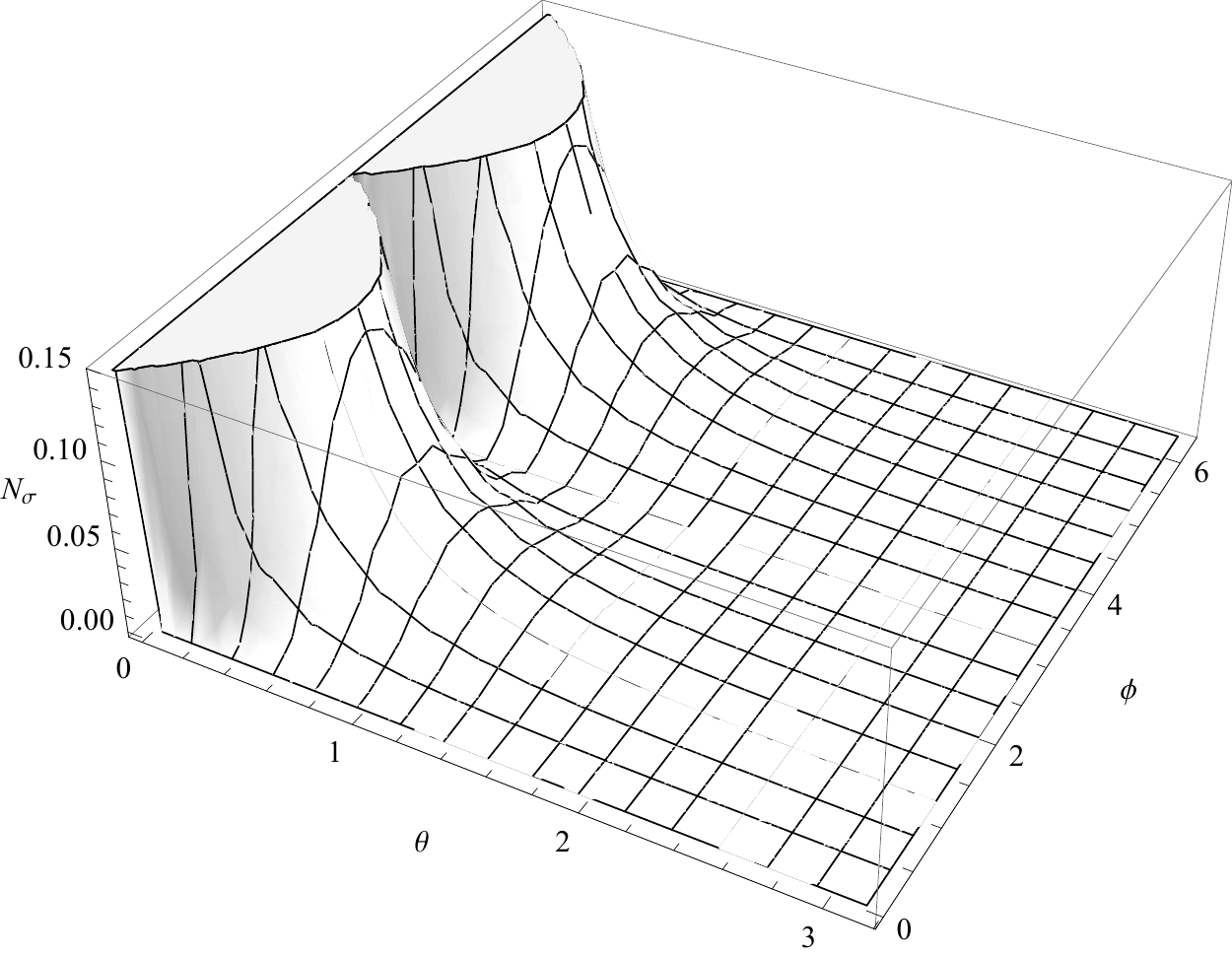} 
\caption{Instantaneous LV differential cross sections for pure space-like background ($\boldsymbol{\xi} \perp \hat{\textbf{z}}$, i.e,. $\theta_\xi = \pi/2$). The vertical axes are given by $N_\sigma = \left[\alpha |\boldsymbol{\xi}|^2 \right]^{-1} d \sigma^{ee, \, \perp}_{_{\rm LV}}/d \Omega$ with $\phi_\xi = 0$ (bottom) and $\phi_\xi = \pi/2$ (top).}
\label{bhabha_fig2}
\end{figure}

In the present case the analysis is slightly more involved due to the amount of angular parameters, hence we focus on two particularizations in order to illustrate the effect of the LV terms. First, let us take a background vector parallel to the $z$-axis ($\theta_\xi = 0$), for which the total differential cross section is given by
\begin{align}\label{cs_b_par}
\frac{d {\sigma}^{ee, \, \parallel}}{d \Omega} &= \frac{\alpha^2 (7 + \cos 2\theta )}{16 E^2_{_{\rm CM}} (\cos\theta - 1)^2} + \\ 
&+\frac{\alpha \, |\boldsymbol{\xi}|^2    ( 17 \cos\theta + 2\cos 2\theta  - \cos 3\theta + 46 ) \sin^4\frac{\theta}{2} }{ 32 \pi (\cos \theta -1)^2}.
\nonumber
\end{align}
Second, we consider a background vector in the transverse $xy$-plane ($\theta_\xi = \pi/2$), where
\begin{equation}\label{cs_b_per}
\begin{split}
\frac{d {\sigma}^{ee, \, \perp}}{d \Omega} & =  \frac{\alpha^2 (7 + \cos 2\theta )}{16 E^2_{_{\rm CM}} (\cos\theta - 1)^2} + \\ 
& +  \frac{\alpha \, |\boldsymbol{\xi}|^2  ( 17 \cos \theta +2 \cos 2\theta - \cos 3\theta + 46 )}{32\pi(\cos \theta -1)^2} \\
&+ \frac{\cos^2(\phi - \phi_{\xi}) \sin^2\theta }{ 32 \pi (\cos \theta -1)^2},
\end{split}
\end{equation}
whose LV piece is plotted in fig.\eqref{bhabha_fig2} for two different choices of the azimuthal angle, $\phi_\xi$. 


From eqs.\eqref{cs_b_par} and \eqref{cs_b_per} above it is clear that the extra LV contribution up to $\mathcal{O}(\xi^2)$ to the total differential cross sections is energy independent, while the pure QED result falls with $E^{-2}_{_{\rm CM}}$. For this reason, with experiments performed at increasingly higher energies, the differential cross section $\frac{d\sigma^{ee}}{d\Omega}$ should, in principle, display an unexpected {\it plateau} for fixed and preferably small scattering angles -- this may be hard to observe experimentally, however. Also, the energies necessary to make this plateau visible would likely be beyond the validity domain of the effective treatment we adopt.

Furthermore, one also notices that there is no resulting azimuthal dependence in the case of a background parallel to the beam axis (cf. fig.\eqref{bhabha_fig2}), whereas the transverse case is clearly $\phi$-dependent; this feature is very distinctive in comparison with the QED result and could, in principle, be visible in high-energy collision experiments.

\section{Pair annihilation} \label{sec_ee}

\indent 

Electron-positron ($e^{-}e^{+}$) annihilation may have several different final states \cite{Marshall,Maluf2}, e.g. $e^{-}e^{+}$, $\mu^{-}\mu^{+}$, $k \, \gamma$ ($k > 1$), etc, but here we are interested in the latter case with $k = 2$, which is the dominating channel with photons in the final state. The practical importance of this reaction lies in the fact that, in $e^{-}e^{+}$ colliders, it represents a large source of background, as it has no lower energy threshold, unlike $e^{-} \, + \, e^{+} \rightarrow f^{-} \, + \, f^{+}$, with $f = e, \mu, \tau, q,\cdots$. A thorough understanding of its features is fundamental to produce precision measurements and to correctly discriminate possible new physics.

This process has also been used as an important tool to study the electroweak interactions between leptons and quarks exchanging $\gamma$ or $Z^0$ bosons in large experiments, such as PETRA, PEP and LEP. Here, however, we limit ourselves to pure QED + LV effects, not taking the full electroweak contributions due to $Z^0$ exchange into account -- this is a safe assumption, since we are aiming at CM energies $\sqrt{s} = 29$ GeV $< m_{_{Z^0}}$, where $\gamma - Z^0$ interference may be neglected \cite{Derrick}.

Below we present two discussions: the calculation of the LV-modified unpolarized differential cross section for $e^{-}e^{+}$ annihilation in two photons and the LV correction to the decay rate of para-positronium.


\subsection{Unpolarized differential cross section}

\indent

The production of two photons via pair annihilation, $e^{-}(p_1) \, + e^{+}(p_2) \rightarrow \gamma(p_3) \, + \, \gamma(p_4)$, is represented at tree level by the $t$- and $u$-channel Feynman diagrams displayed in fig.\eqref{fig:diagrams} and here we are interested in the unpolarized differential cross section, i.e., we do not keep track of spin orientations and polarizations. In standard QED, it is found that
\begin{equation}
\frac{d\sigma^{\gamma\gamma}_{_{\rm QED}}}{d\Omega} = \frac{\alpha^2}{2E^2_{_{\rm CM}}}\frac{1 + \cos^2\theta}{\sin^2\theta}, \label{ann_qed}
\end{equation}
where the ultra-relativistic limit is assumed: $E = E_{_{\rm CM}}/2 \simeq |\textbf{p}| \gg m$. The process is evaluated in the CM frame, where $p_1 = (E,\textbf{p})$ and $p_2 = (E,-\textbf{p})$ with $\textbf{p} = E\hat{\textbf{z}}$, while $p_3 = (E,\textbf{k})$ and $p_4 = (E,-\textbf{k})$, with $|\textbf{k}| = E$. We have introduced a symmetry factor $S = 1/2$ to account for the identical particles in the final state.

As in section \ref{sec_bhabha}, we are interested in obtaining the total differential cross section for this process with the modified vertex, eq.\eqref{vertex}, so that a comparison with experimental limits may lead to bounds on the LV parameters. The detailed calculation of the squared amplitude for this process and its subsequent averaging is a lengthy and cumbersome task, but it can be greatly simplified by realizing that the amplitude for pair annihilation is connected to that of Compton scattering through crossing symmetry.

Following this procedure we find that, as with Compton scattering -- see eq.\eqref{KN_lv} -- the squared amplitude for $e^{-}e^{+}$ annihilation brings the LV effects as a pre-factor of $(\xi \cdot p_3)^2 + (\xi\cdot p_4)^2$, so that, applying the kinematics mentioned above, we find that this factor becomes 
\begin{equation}
(\xi\cdot p_3)^2 + (\xi\cdot p_4)^2 = \frac{E^2_{_{\rm CM}} }{2} \left[ \xi_0^2 + |\boldsymbol{\xi}|^2 a^2(\theta,\phi,\theta_\xi,\phi_\xi)  \right], \label{nn}
\end{equation}
where $a(\theta,\phi,\theta_\xi,\phi_\xi) = \sin\theta \sin\theta_\xi \cos(\phi - \phi_\xi) + \cos\theta\cos\theta_\xi$.

The (instantaneous) total differential cross section in the high-energy limit, up to $\mathcal{O}(\xi^2)$, can then be conveniently expressed as
\begin{align}
\frac{d\sigma^{\gamma\gamma}}{d\Omega} &= \frac{\alpha^2}{2E^2_{_{\rm CM}}}\frac{1 + \cos^2\theta}{\sin^2\theta} \times \nonumber \\
&\times \Bigg\{ 1 + \frac{E^2_{_{\rm CM}}}{16\pi\alpha} \left[ \xi_0^2 + |\boldsymbol{\xi}|^2 a^2(\theta,\phi,\theta_\xi,\phi_\xi)  \right]  \Bigg\}, \label{ann_LV}
\end{align}
and we notice that, as in the case of Bhabha scattering, the LV contribution is overall energy-independent. An analysis of the angular dependence of the pure LV piece of eq.\eqref{ann_LV} would lead to conclusions similar to those obtained in the previous sections: for a space-like background aligned with the initial electron-positron motion, there is no $\phi$-dependence -- only an extra $\sim \cos^2\theta$ factor is added, while for a transverse background, similar peaks as those depicted in fig.\eqref{fig1} are expected. We note, furthermore, that, in this latter configuration, the forward peak ($\theta \rightarrow 0$) is absent due to the additional $\sin^2\theta$ factor from eq.\eqref{nn}.

We are now ready to compare eq.\eqref{ann_LV} with eq.\eqref{ann_qed} in a more concrete way. Time-averaged deviations from the standard QED tree-level prediction for $e^{-}e^{+}$ annihilation are bounded by experiment via 
\begin{equation}
\Bigg| \frac{d {\sigma}^{\gamma\gamma, \, \xi_0}/d \Omega}{d \sigma^{\gamma\gamma}_{_{\rm QED}}/d \Omega} - 1 \Bigg| \lesssim \frac{ E_{_{\rm CM}}^4}{2\tilde{\Lambda}^4}, \label{criterio_2}
\end{equation}
at $95 \%$ CL with $\tilde{\Lambda} = 59 \, {\rm GeV}$ at $\sqrt{s} = 29 \, {\rm GeV}$ \cite{Derrick}. Using this constraint with eq.\eqref{ann_LV} we arrive at
\begin{equation}
\xi_0 \lesssim 10^{-3} \, {\rm GeV}^{-1}  \label{limit_ee}
\end{equation}
as an upper bound on the LV parameters.



\subsection{Life-time of para-positronium} \label{posit}

\indent 

Positronium is the bound state of an electron and a positron. It was predicted in the 1930's and experimentally observed by Deutsch in 1951 \cite{Deutsch} and its main decay channels are in two or three photons for the singlet (para-positronium, p-Ps) and triplet (ortho-positronium, o-Ps) spin states, respectively \cite{Savely}. Here we shall focus on the LV contribution to the life-time of p-Ps, given in QED by the inverse of the decay rate, $\Gamma_{2\gamma, \, QED} = \frac{m\alpha^5}{2}$. Its experimental value, which agrees well with theory \cite{Melnik, Penin}, was measured to be 125 ps, with a relative precision of $215 \, {\rm ppm}$, that is $\delta_{\tau} \sim 10^{-4}$ \cite{life_time}.


The decay rate of o-Ps is not significantly more precise than that of p-Ps and its relative precision reads 150 ppm \cite{Asai}. The former is, however, a higher order process in QED, so we go for the simplest one, p-Ps, without significant loss regarding the outcoming bound on the LV parameters.

The decay rate of p-Ps in two photons, although closely related to the calculation performed above, does not follow as a direct sub-product of the previous result.
As a matter of fact, when computing the cross section for pair annihilation we were interested in the unpolarized result in the ultra-relativistic limit, whereas in the present case we consider
that the kinetic energies of both the electron and the positron are much smaller than their rest energies. We also take the spin-polarized case of the singlet state.

Imposing the appropriate limit, the correct polarization state and only keeping the lowest order contribution in the Lorentz-violating
coupling, the squared amplitude is only sensitive to the time component of the background vector. Moreover, the result is now isotropic and there is 
no interference between the pure QED and the LV sectors:
\begin{equation}
|\mathcal{M}_{\rm tot}|^2 = |\mathcal{M}_{\rm QED}|^2 + |\mathcal{M}_{\rm LV}|^2 = 16e^4 \left(1 + 4\frac{m^2 \xi^2_0}{e^2} \right). \label{amp_pos}
\end{equation}

Therefore, the total decay rate, including the well-known QED term and the extra LV contribution, is then
\begin{equation}
\Gamma_{2\gamma} = \frac{m\alpha^5}{2} \left(1 + \frac{m^2 \xi^2_0 }{\pi \alpha} \right), \label{life_time}
\end{equation}
and, assuming the LV part to be very small ($\xi^2_0 \ll 1/m^2$), we may write the modified life-time of p-Ps to a good approximation as $\tau_{_{2\gamma}} \equiv \Gamma_{2\gamma}^{-1} \simeq \frac{2}{m\alpha^5} \left(1 - \frac{m^2 \xi^2_0}{\pi \alpha} \right)$. Demanding that the experimental result is well fit by pure QED, we may set an upper limit on the LV parameters by requiring that the associated LV correction does not extrapolate the experimental error, i.e.,
\begin{equation}
\Bigg| \frac{ \tau_{_{2\gamma}} }{\tau_{_{2\gamma , \, {\rm QED} }}} - 1 \Bigg|  \lesssim \delta_{\tau}. \label{criterio_3}
\end{equation}
Applying this criterium, we find the upper bound:
\begin{equation}
\xi_0 \lesssim 1 \, {\rm GeV}^{-1}. \label{bound_pos}
\end{equation}

It is important to highlight that other authors have considered the effects of different LV sectors in positronium (e.g. ref. \cite{vargas} and ref. \cite{adkins}). In this context, the results for spectroscopy measurements are specially interesting, for these experiments are extremely precise, what could give more restrictive bounds on the LV parameters.


\section{Concluding remarks} \label{sec_conclusion}

\indent

In this paper we have discussed the modifications in simple QED processes due to the inclusion of a new non-minimal coupling between the electron and the photon \cite{bakke}, eq.\eqref{case_1}. We found that novel energy- and angle-dependent corrections arise already at lowest order in the LV parameter and, up to this order, we were able to establish upper limits on $\xi_0$ by demanding that the LV-modified physics do not exceed the established QED results by more than a few percent -- see eqs.\eqref{criterio_1}, \eqref{criterio_2} and eq.\eqref{criterio_3}. Similar limits on $|\boldsymbol{\xi}|$ should be expected, cf. section \ref{sec_intro}.


The limit from p-Ps owes its relative weakness to the fact that, contrary to the other processes studied, it is a non-relativistic system, since the initial $e^{-} e^{+}$ pair is taken as being practically at rest. 
The LV non-minimal coupling we consider brings an energy-dependent correction, which, in the low-energy limit ($\textbf{p} \simeq 0$ and $E \simeq m$), means that the LV correction to the decay rate becomes $|\mathcal{M}_{_{\rm LV}}|^2\sim \xi_0^2 E^2 \rightarrow \xi_0^2 m^2$, which is also expected on purely dimensional grounds. These considerations, combined with the relatively large uncertainty ($\delta_{\tau} = 215 \, {\rm ppm}$), are responsible for the looser bound quoted in eq.\eqref{bound_pos}. As mentioned in the end of section \ref{posit}, an application of our modified vertex to the spectra of simple atoms (hydrogen, positronium, etc) may improve the limits quite significantly, as spectroscopic measurements reach uncertainties as low as $10^{-15}$ \cite{Colladay3, Parthey}.

We have focused on purely time- or space-like background configurations, as is customary in the field of LV, but it is clear that such a division is arbitrary -- if such a background exists, it would likely be a non-trivial mixture of such components. However, since we are interested in estimating upper bounds for the background $\xi$, we refrain from stating a more general result taking a background with time- and space-like components, as this would not improve neither the readability of the results nor the bounds obtained.

It is worthwhile pointing out that o-Ps, being the triplet spin state of positronium, naturally offers a 3-vector (the polarization) that should couple to the external background. This means that an analysis of o-Ps could bring information on the spatial components of the background. However, since the experimental uncertainty is of similar size as for p-Ps \cite{Asai}, we do not expect better limits.

We would like to point out that, even though p-Ps may be used to extract upper limits on the background, it is probably does not provide a `smoking gun' for LV, since other beyond-the-Standard-Model scenarios could possibly generate similar effects. In this context, it is possible that o-Ps, being a triplet state, may be directionally more sensitive to a fixed spatial background, thus more sensitive to sidereal variations, which could provide an unambiguous signal of LV.

We have found that, for a pure space-like background, the instantaneous LV-modified differential cross sections generally present periodic contours as a function of the azimuthal angle -- see e.g. figs.\eqref{fig1} and \eqref{bhabha_fig2}. The respective pure QED processes do not discriminate different $\phi$-orientations and this is a clear LV signal that could be searched for in collider experiments. Another interesting feature is the scaling of the LV contributions with energy, which enhances its effects in high-energy experiments (in contrast to the QED contributions), possibly allowing for future direct tests of LV in accelerators.

Ignoring the boost factors as above, we note that the temporal and spatial components do not mix, cf. section \ref{sec_intro}. This implies that our independent analysis of pure time- and space-like backgrounds may be extended to the components in the SCF. For the case where $\xi^{J}_{\rm sun} \equiv 0$, the upper bounds from eqs.\eqref{limit_1} and \eqref{limit_ee} immediately translate to
\begin{equation}
\xi_{\rm sun}^{T} \lesssim 10^{-3} \, {\rm GeV}^{-1}
\end{equation}
at $95 \, \%$ CL, whereas from eq.\eqref{bound_pos} we get 
$\xi_{\rm sun}^{T} \lesssim 1 \, {\rm GeV}^{-1}$. Despite of the relative simplicity of the pure time-like case, the pure space-like scenario ($\xi^{T}_{\rm sun} \equiv 0$) is significantly more cumbersome due to the contractions with the 3-momenta of the participating particles. For this reason we choose not to display the result in terms of the SCF variables, as no new significant physical information would be conveyed.


Finally, we would like to indicate that other sectors from the SME \cite{Colladay, Colladay2} could also induce effects in scattering processes, e.g. the $k_{\rm AF}$ and $k_{\rm F}$ contributions to the photon sector. Since both enter in the quadratic part of the Maxwell Lagrangian, they would modify the photon propagator, also demanding corrections to the dispersion relations and polarization 4-vectors (see e.g. \cite{Schreck}). These corrections would also be momentum-dependent and would potentially lead to modifications in the (differential) cross sections in QED processes, including the ones treated above. While Schreck \cite{Schreck} has already discussed the $k_{\rm F}$ sector in connection to Compton scattering, we think that a similar analysis of the Carroll-Field-Jackiw $k_{\rm AF}$ term \cite{CFJ} would be worthwhile, as it could provide complementary local, i.e., non-astrophysical, bounds on this LV parameter.

Besides the points addressed in this paper, it is also possible that the class of non-minimal couplings studied here -- also with axial couplings containing $\gamma_5$ -- could produce interesting contributions to the $g-2$ of the electron (or more interestingly, of the muon) already at tree-level \cite{Araujo}. The tight experimental constraints on $a_{e,\mu}= (g-2)/2$ together with the expected momentum dependence of the extra LV vertices should allow for better upper limits on $\xi^{\mu}$. The task of quantitatively evaluating this prospect is currently being undertaken and shall be presented elsewhere.


\begin{acknowledgments}
The authors are grateful to J.A. Helay\"el-Neto, J. Jaeckel and V.A. Kosteleck\'y for interesting discussions and for reading the manuscript. This work was funded by the Brazilian National Council for Scientific and Technological Development (CNPq) and the German Service for Academic Exchange (DAAD). P.C.M. would also like to thank the Institut f\"ur theoretische Physik (U. Heidelberg) for the hospitality during the elaboration of this work.
\end{acknowledgments}



\end{document}